\documentclass{Interspeech2024}

\usepackage{multirow}
\usepackage{pifont}
\newcommand{\cmark}{\ding{51}}%
\newcommand{\xmark}{\ding{55}}%
\usepackage[dvipsnames]{xcolor}

\interspeechcameraready

\title{SA-WavLM: Speaker-Aware Self-Supervised Pre-training for Mixture Speech\vspace{-0.7cm}}

\name[affiliation={1}]{Jingru}{Lin}
\name[affiliation={1,*}]{Meng}{Ge}
\name[affiliation={3}]{Junyi}{Ao}
\name[affiliation={2}]{Liqun}{Deng}
\name[affiliation={1,3}]{Haizhou}{Li}

\address{
  \vspace{-0.25cm}
  $^1$Department of Electrical and Computer Engineering, National University of Singapore, Singapore\\
  $^2$ Huawei Noah's Ark Lab \quad $^3$School of Data Science, Shenzhen Research Institute of Big Data, The Chinese University of Hong Kong, Shenzhen, China 
  \vspace{-0.12cm}
  }
\email{jingrulin@u.nus.edu, 
\{gemeng, haizhou.li\}@nus.edu.sg
\vspace{-1.5cm}}


\begin{document}

\maketitle

\begin{abstract}
It was shown that pre-trained models with self-supervised learning (SSL) techniques are effective in various downstream speech tasks. 
However, most such models are trained on single-speaker speech data, limiting their effectiveness in mixture speech. This motivates us to explore pre-training on mixture speech. This work presents SA-WavLM, a novel pre-trained model
for mixture speech. 
Specifically, SA-WavLM follows an ``extract-merge-predict'' pipeline in which the representations of each speaker in the input mixture are first extracted individually and then merged before the final prediction. In this pipeline, SA-WavLM performs speaker-informed extractions with the consideration of the interactions between different speakers. Furthermore, a speaker shuffling strategy is proposed to enhance the robustness towards the speaker absence. 
Experiments show that SA-WavLM either matches or improves upon the state-of-the-art pre-trained models.

\end{abstract}

\renewcommand{\thefootnote}{}
\footnotetext{This work is supported by 1) Huawei Noah's Ark Lab; 2) Shenzhen Science and Technology Research Fund (Fundamental Research Key Project Grant No. JCYJ20220818103001002); 3) Shenzhen Science and Technology Program ZDSYS20230626091302006; 4) National Research Foundation Singapore under its AI Singapore Programme grant number AISG2-TC-2022-004. $^*$ Corresponding author.}

\section{Introduction}
Self-supervised learning (SSL) based pre-training  
have greatly advanced over the past few years~\cite{mohamed2022self,ericsson2022self,NEURIPS2023_9d276b0a}. By leveraging on a large amount of unlabeled speech data, the pre-trained models can produce universal speech representations that benefit a wide range of downstream applications, such as speech recognition, speaker verification, acoustic word embeddings etc~\cite{schneider19_interspeech,chen2022large,lin23d_interspeech}. However, many existing pre-trained models, e.g., wav2vec 2.0~\cite{baevski2020wav2vec} and HuBERT~\cite{hsu2021hubert}, are designed for clean speech, limiting their effectiveness in handling the mixture speech. 

Mixture speech, where multiple speakers can speak simultaneously, presents a notably challenging scenario.
WavLM~\cite{chen2022wavlm} emerged as an early attempt to address this limitation. 
Given simulated mixture speech, WavLM is expected to perform denoising and masked prediction of the clean speech. A constraint of mixing portion to be less than 50\% is imposed during mixture simulation, and a primary speaker is defined based on a longer speech duration. During pre-training, WavLM focuses only on masked prediction of the primary speaker's speech while disregarding other interfering speakers.
Similar approaches, such as those seen in~\cite{zhu2023robust,wang2022wav2vec}, define the primary speaker as the solo speaker in non-speech background noises. 
Consequently, the resulting representations contain only information about a single speaker. Although these strategies help pre-trained models generalize better to mixture speech, subsequent research has suggested that these pre-trained models are tailored for single-speaker speech and may be suboptimal for tasks involving mixture speech~\cite{fazel2023cocktail}, where all speakers are equally important.

The capability to process mixture speech is crucial for the well-known cocktail party problem~\cite{qian2018past,cherry1953some}. It is therefore important to design pre-trained models to handle mixture speech. Recently, several studies have attempted to extend the applicability of pre-trained models to mixture speech. In~\cite{wang2023adapter}, both single-speaker and two-speaker mixture speech are simulated as input during the pre-training, and the pre-trained model learns to predict the masked timestamps of all clean speeches given the input mixture speech. A similar training strategy is extended to mixture speech with even more speakers in~\cite{fazel2023cocktail}, by using a permutation invariant training (PIT) loss~\cite{yu2017permutation}. Both studies devise a pre-training scheme to model the mixture representations given the mixture input. As they are concerned about all speakers in the mixture speech, the pre-trained model's ability to handle mixture speech is greatly enhanced. 

While the above-mentioned methods are effective, this work explores another promising direction: leveraging additional speaker information. The use of speaker information has show efficacy in various tasks, including speaker-attributed automatic speech recognition (ASR)~\cite{kanda21b_interspeech,kanda2022transcribe} and 
speaker diarization (SD)
tasks~\cite{medennikov20_interspeech}. 
In~\cite{kanda21b_interspeech}, speech recognition and speaker identification are jointly performed for multi-speaker speech scenarios. In~\cite{medennikov20_interspeech}, the speech activities of all speakers in the input mixture speech are estimated given the speaker profile of each speaker. These methods demonstrate the effectiveness of integrating the speaker information into the model for addressing mixture speech challenges. 


\begin{figure*}[t]
  \centering
  \includegraphics[width=0.85\linewidth]{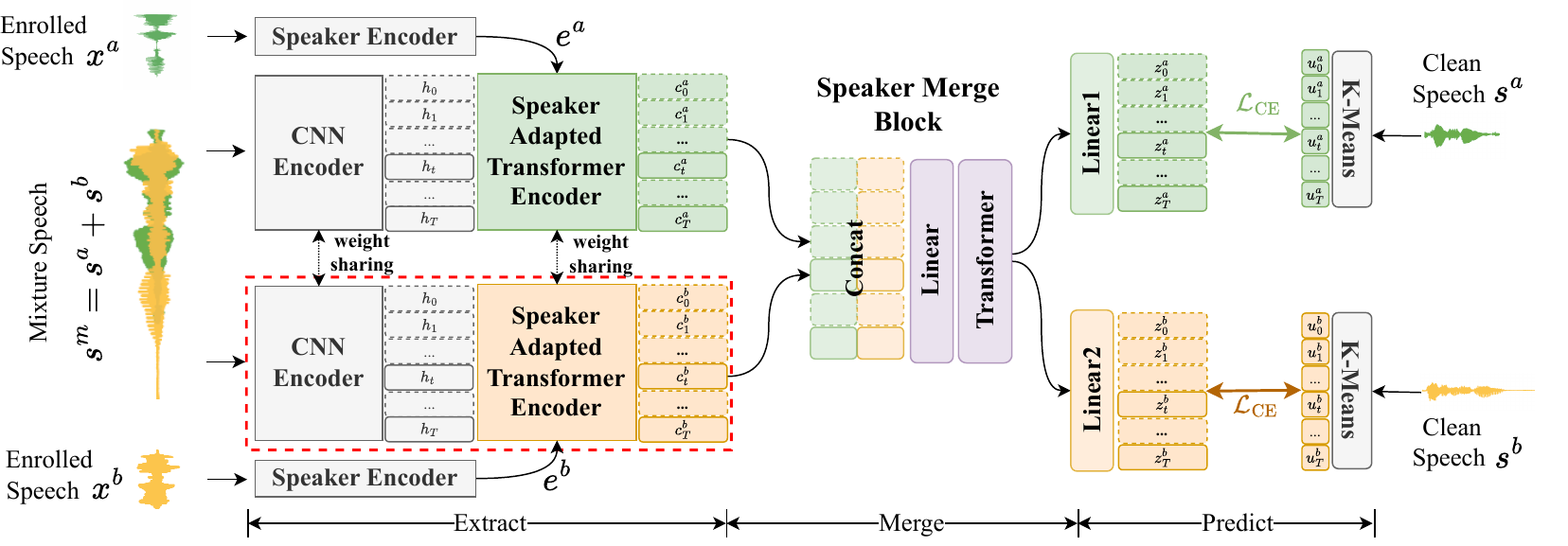}
  \caption{The overview of SA-WavLM architecture and the ``extract-merge-predict" pipeline. Given the input mixture speech, the proposed model first extracts the individual representations for each speaker in the input. Subsequently, the Speaker Merge Block merges the individual representations and models their interactions before making the final prediction. 
  In downstream tasks, only the ``extract'' stage is used, which is the part within the red dotted box.
  }
  \label{fig:architecture}
  \vspace{-0.4cm}
\end{figure*}

Inspired by the ideas discussed above, we propose SA-WavLM, a novel speaker-aware self-supervised pre-trained model designed to better handle the mixture speech. 
In contrast to WavLM~\cite{chen2022wavlm}, which focuses solely on a primary speaker that might neglect others, our SA-WavLM addresses all speakers within the input mixture speech. To achieve this, SA-WavLM adopts an ``extract-merge-predict'' pre-training pipeline. Beginning with the ``extract'' phrase, SA-WavLM obtains individual representations of each speaker in the mixture speech, conditioned on the corresponding speaker information. The speaker's information is provided by speaker embeddings, which are injected into SA-WavLM by conditional layer normalization.
Next, in the ``merge'' phrase, SA-WavLM combines the individual representations and fosters interactions between different speakers via a Speaker Merge Block. Finally, SA-WavLM will ``predict'' pseudo labels for each speaker. Furthermore, we introduce a speaker shuffling strategy to ensure SA-WavLM's invariance to speaker order and absence. This pre-training pipeline eliminates the need for constraints on simulated mixture data, as seen in WavLM, and empowers SA-WavLM with enhanced interference removal ability, even in scenarios where the target speaker is non-primary in the mixture.


\vspace{-0.2cm}
\section{WavLM}
\vspace{-0.1cm}
Our work is most related to WavLM~\cite{chen2022wavlm}, a self-supervised speech pre-trained model that combines denoising and masked speech prediction during pre-training. WavLM consists of a convolutional neural network (CNN) $\mathcal{F}(\cdot)$ and a Transformer encoder $\mathcal{G}(\cdot)$.
Given a mixture speech $s^m$ simulated from the clean speeches $s^a$ of speaker $a$ and $s^b$ of speaker $b$, WavLM first extracts the frame-level speech representations $H^m = \left[h^m_1, h^m_2, ..., h^m_T\right]$ through $\mathcal{F}(\cdot)$. These representations are the passed through a partial masking process $\mathcal{M}(\cdot)$ and fed into $\mathcal{G}(\cdot)$. 
Let us assume speaker $a$ is the primary speaker whose speech $s^a$ is longer than $s^b$. 
WavLM performs denoising for speaker $a$ to get the contextual representations $C^a = \left[c^a_1, c^a_2, ..., c^a_T\right]$ that corresponds to speech $s^a$.
The whole pipeline can be formalized as 
\vspace{-0.1cm}
\begin{equation}
    C^a = \left[c^a_1, c^a_2, ..., c^a_T\right] = \mathcal{G}\left[\mathcal{M}(\mathcal{F}(s^m))\right]
    \vspace{-0.1cm} 
    \label{eqn:wavlm}
\end{equation}

In the masked prediction process, the generated $C^a$ should be able 
to predict the pseudo labels $U^a = \left[u^a_1, u^a_2, ..., u^a_T\right]$ for the masked frames in $s^a$. The masked speech prediction loss during the pre-training is the cross-entropy (CE)
\vspace{-0.1cm}
\begin{equation}
    \mathcal{L}_m = \mathcal{L}_{\text{CE}}(C^a, U^a) =  \sum\nolimits_{t \in O} \log \, p_t (\, u^a_t \, | \, c^a_t)
    \vspace{-0.1cm}
\end{equation}
where $O$ denotes the set of masked indices in $H^m$. An offline clustering model (i.e. k-means) is used to get the pseudo labels. 

In Eq. (\ref{eqn:wavlm}), $s^a$ is assumed to have a longer speech duration in $s^m$. In fact, during the mixture simulation for pre-training WavLM, a constraint is applied such that the mixing portion must be less than 50\%. This constraint ensures that the speech from the primary speaker is always longer than the other speech, thereby informing the model about the primary speaker (to retain) and the interfering speaker (to remove).
In practice, alternative primary speaker definitions, such as the speaker with higher energy or the solo speaker in noisy speech (clean speech with background non-speech noises), are viable too.

\vspace{-0.2cm}
\section{Proposed SA-WavLM}
In WavLM, only the primary speaker is concerned. However, the definition of a primary speaker may not align with multi-speaker speech tasks, including speech diarization, speech separation, and multi-speaker speech recognition, where all speakers in the mixture speech are equally important. 

Building upon WavLM, we propose SA-WavLM to address this limitation. 
Figure~\ref{fig:architecture} illustrates the overview of SA-WavLM architecture and the ``extract-merge-predict'' pipeline. 
Given the mixture speech $s^m = s^a + s^b$ 
, where $s^a$ and $s^b$ are two clean speeches from speakers $a$ and $b$, we first ``extract'' representations of speakers $a$ and $b$ individually by the Speaker Adapated Transformer Encoder (SATE) denoted as $\mathcal{G_{\text{SATE}}(\cdot)}$, replacing the $\mathcal{G}(\cdot)$ in WavLM. The individual representations are then ``merged'' using a Speaker Merge Block (SMB), which facilitates their interactions. Finally, the training objective of SA-WavLM is to ``predict'' the pseudo labels of speeches $s^a$ and $s^b$. In addition,  to ensure the model's invariance towards speaker order and absence, we introduce a speaker shuffling strategy.


\vspace{-0.2cm}
\subsection{Speaker Adapted Transformer Encoder (extract)}
\vspace{-0.1cm}
Speaker Adapted Transformer Encoder (SATE) is designed to extract the contextual representation for the target speaker. Given speaker embedding $e^k$ of the target speaker $k \in \{a,b\}$ and masked frame-level representation $H^m = \left[h^m_1, h^m_2, ..., h^m_T\right]$ of $s^m$ extracted by $\mathcal{F}(\cdot)$, SATE, denoted by $\mathcal{G_{\text{SATE}}}$, is trained to extract the contextual representation $C^k = \left[c^k_1, c^k_2, ..., c^k_T \right]$ that corresponds to $s^k$. This is formulated as
\vspace{-0.1cm}
\begin{equation}
    C^k = \left[c^k_1, c^k_2, ..., c^k_T\right] = \mathcal{G_{\text{SATE}}}\left[\mathcal{M}(\mathcal{F}(s^m)), e^k\right], k \in \{a,b\}
    \label{equation:SATE}
    \vspace{-0.1cm}
\end{equation}

SATE consists of a Speaker Adapated Transformer Layer (SATL) and multiple Vanilla Transformer Layers (VTLs). Here the VTL in SATE has the same structure as the Transformer layer in the Transformer encoder $\mathcal{G}(\cdot)$ of WavLM. 
The main extraction process of SATE is performed through SATL, which replaces the traditional layer normalization in VTL with conditional layer normalization (CLN). In VTL, the layer normalization operation is given by
\vspace{-0.1cm}
\begin{equation}
    H_{\text{VTL}} = \frac{H^{m} - E\left[H^{m}\right]}{\sqrt{Var\left[H^{m}\right] + \epsilon}} \otimes \gamma + \beta, 
    \vspace{-0.1cm}
\end{equation}
where $E\left[H^{m}\right]$ and $Var\left[H^{m}\right]$ are the mean and variance of the input representation $H^{m}$, and $\gamma$ and $\beta$ are learnable weight and bias for applying the element-wise affine transformation.
    
To extract the representations of speech $s^k$ given $e^k$, the CLN in SATL replaces $\gamma$ with speaker-specific scaling, i.e.
\vspace{-0.1cm}
\begin{equation}
    H_{\text{SATL}} = \frac{H^{m} - E\left[H^{m}]\right]}{\sqrt{Var[H^{m}] + \epsilon}} \otimes \left[w(e^k) \cdot \gamma + \theta(e^k)\right] + \beta,
\end{equation}
where $w$($\cdot$) and $\theta$($\cdot$) are linear projections to project the speaker embedding $e^k$ to the feature dimension of $H^{m}$ to perform the element-wise multiplication. In fact, CLN modulates the normalization output via a specific speaker's latent representation. 

\vspace{-0.2cm}
\subsection{Speaker Merge Block (merge)}
Motivated by TS-VAD~\cite{medennikov20_interspeech}, it is believed that interaction between different speakers is important for better extraction and separation abilities. Therefore, in SA-WavLM, we design a Speaker Merge Block (SMB) to allow interaction between different speakers. The structure of SMB is also shown in Figure~\ref{fig:architecture}. 

SMB first concatenates $C^a$ and $C^b$ of speaker $a$ and $b$, extracted respectively by SATE as shown in Eq. (\ref{equation:SATE}), along the feature dimension. Later, the concatenated representations are projected down by a linear layer. A single VTL, which has the same structure as the VTL in SATE, is deployed to model the down-projected representations and their interactions, generating the mixture contextual representations $C^m$
\vspace{-0.1cm}
\begin{align}
    C^m = \text{VTL}(\text{Linear}(\text{Concat}(C^a, C^b)))
    \label{equation:SMB}
\end{align}
Note that the SMB is used during pre-training only to facilitate the interactions. In downstream applications, SMB is removed and only CNN and SATE are used. 

\vspace{-0.2cm}
\subsection{Prediction and training objective (predict)}
Unlike Cocktail HuBERT~\cite{fazel2023cocktail} where PIT~\cite{yu2017permutation} is needed to find the optimal alignment between the predictions and pseudo labels, SA-WavLM determines the order of predictions based on the order of speaker embeddings injected. 
That is, if the concatenation order in Eq. (\ref{equation:SMB}) is $a$ followed by $b$, Linear1 predicts the pseudo labels of clean speech $s^a$ and Linear2 predicts that of clean speech $s^b$ and vice versa. Here, the former order, i.e. $a$ followed by $b$, is used for illustration
\vspace{-0.1cm}
\begin{equation}
    Z^a = \text{Linear1}(C^m), \quad Z^b = \text{Linear2}(C^m)
    \vspace{-0.1cm}
\end{equation}

The final masked speech prediction loss is the summation of Cross Entropy (CE) losses for the two speakers 
\vspace{-0.1cm}
\begin{equation}
    \mathcal{L}_m  = \sum\limits_{k \in \{a,b\}} \mathcal{L}_{\text{CE}}(Z^k, U^k)
    = \sum\limits_{k \in \{a,b\}} \sum\limits_{t \in O^k} \log \, p_t (\, u^k_t \, | \, z^k_t)
    \vspace{-0.1cm}
\end{equation}

\vspace{-0.3cm}
\subsection{Speaker shuffling strategy}
To keep the model invariant to speaker order, we deploy a speaker shuffling strategy to shuffle the order of speaker embeddings and the corresponding pseudo labels. In our mixture simulation, we have two-speaker scenarios and one-speaker scenarios. Two-speaker scenarios include two-speaker overlapped speech $s^m=s^a + s^b$ and noisy two-speaker overlapped speech $s^m=s^a + s^b + n$, where $n$ is the background noise. One-speaker scenarios include clean speech $s^a$ and noisy single-speaker speech $s^n=s^a+n$. 
For two-speaker scenarios, the speaker embeddings $e^a$ and $e^b$ are used directly. While for one-speaker scenario, we will use speaker embeddings $e^a$ and, either 1) $e^c$ from a random speaker that is different from speaker $a$, with a probability of $\alpha$, or 2) $e^s$, a learnable vector indicating non-speaker existence, with a probability of $(1-\alpha)$. Subsequently, we shuffle the order of all the speaker embeddings injected into SATE, which determines the order of the Concat operation in Equation \ref{equation:SMB}. This strategy aims to enhance the model's insensitivity towards the speaker order and robustness towards the speaker absence, thereby improving the accuracy of extraction. The pseudo labels corresponding to $e^c$ or $e^s$ will be $S$, which is a vector of silence tokens.

\vspace{-0.2cm}
\section{Experimental setups}

\vspace{-0.1cm}
\subsection{Training datasets}
\vspace{-0.1cm}
Following~\cite{wang2023adapter}, we pre-train our model on the data simulated from the 960-hour LibrSpeech~\cite{panayotov2015librispeech} and the DNS Challenge's noise datasets~\cite{reddy20_interspeech}.
Due to limited computational resources, only one- and two-speaker scenarios are simulated. This includes clean speech, noisy single-speaker speech, two-speaker overlapped speech, and noisy two-speaker overlapped speech, which are generated through a dynamic mixing strategy~\cite{zhang2023weakly}. The enrolled speech of the target speaker is randomly selected from the LibriSpeech corpus, and it should be different from the one used to simulate the mixture speech. The speaker embedding is generated from the selected enrolled speech using the pre-trained CAM++~\cite{wang2023cam++}. All speeches are sampled at 16kHz.

\vspace{-0.2cm}
\subsection{Models and training details}
\vspace{-0.2cm}
All pre-training experiments are conducted using Fairseq toolkit~\cite{ott-etal-2019-fairseq}. Our model is based on WavLM Base~\cite{chen2022wavlm}, while replacing a Vanilla Transformer Layer (VTL) with a Speaker Adapted Transformer Layer (SATL) and adding a Speaker Merge Block (SMB). SATL can replace any VTL in the Transformer encoder, but here we only replace the first layer.
The parameters of CNN encoder and SATE are initialized from the pre-trained WavLM Base, 
except for $w$ and $\theta$ of SATL. For the SMB, the weights are randomly initialized. Training takes 400k steps with a learning rate of 7e-5, using the pseudo labels generated from the 9th transformer layer of the HuBERT Base model~\cite{fazel2023cocktail}. The probability $\alpha$ used in the speaker shuffling strategy is set to 0.5. Other hyperparameters are consistent with those of the WavLM Base. 
Note that although SA-WavLM has the largest model size as shown in Table~\ref{tbl:superb}, the additional SMB is removed for downstream tasks. This leaves our model size with 94.97M parameters, which is comparable to other pre-trained models (e.g. WavLM Base, HuBERT Base etc).

\vspace{-0.3cm}
\subsection{Downstream evaluations}
\vspace{-0.2cm}
To evaluate the effectiveness of our models on the cocktail party problem, we evaluated our models on different mixture speech tasks including speech enhancement (SE), separation (SS), diarization (SD), extraction (Ext) and multi-speaker speech recognition (ASR). For SE, the dataset used is Voicebank-DEMAND~\cite{veaux2013voice}. For the remaining, the different subsets of Libri2Mix\footnote{Available at https://github.com/JorisCos/LibriMix}
are used. In particular, SD uses the ``mix-both max mode'', ASR uses the ``mix-clean max mode'' and SS and Ext use the ``mix-clean min mode''. 
All the datasets in downstream evaluations have 16kHz sample rates.

For the baselines, we select the state-of-the-art pre-trained models including wav2vec 2.0~\cite{baevski2020wav2vec}, HuBERT~\cite{hsu2021hubert}, WavLM ~\cite{chen2022wavlm}, Wang et.al~\cite{wang2023adapter} and Cocktail HuBERT~\cite{fazel2023cocktail}. For all the baselines, Base sizes, which are comparable to SA-WavLM, are used. Among them, wav2vec 2.0 and HuBERT are trained on clean speech only, while the rest are trained on mixture speech. 

\vspace{-0.3cm}
\section{Experiments}

\subsection{Results on SUPERB benchmark}
We conduct evaluations on SUPERB~\cite{yang21c_interspeech}, the well-known benchmark that unifies evaluations on various speech processing tasks. It provides lightweight downstream networks that make it easy to compare and draw insights across different pre-trained models. 
Assessments are made for the provided mixture speech tasks: SE, SS and SD.
We report Perceptual evaluation of speech quality (PESQ) and short-time objective intelligibility (STOI) for SE, scale-invariant signal-to-distortion ratio improvement (SI-SDRi) for SS, and the diarization error rate (DER) for SD. In all the experiments, the pre-trained models are frozen and only the downstream networks are fine-tuned.

Table~\ref{tbl:superb} lists the evaluation results on the SUPERB benchmark. It can be observed that wav2vec 2.0 and HuBERT do not generalize well to the mixture speech tasks. With the denoising strategy, WavLM outperforms wav2vec 2.0 and HuBERT across all three tasks. Both Wang et. al and Cocktail HuBERT (referred to as C-HuBERT in Table~\ref{tbl:superb}) are designed for mixture speech, hence demonstrating superior performance on all three tasks. When more speakers are considered during pre-training, C-HuBERT shows further improvement on SS. Compared against all baselines, SA-WavLM either matches or outperforms across all the tasks. Notably, both Wang et.al and SA-WavLM are pre-trained with 2-speaker mixture speech only, yet SA-WavLM demonstrates better performance on SS, surpassing Wang et. al by 0.54dB SI-SDRi. This proves the effectiveness of the use of speaker cues and modeling of speaker interactions. 

\begin{table}[t]
\centering
    \scriptsize
    \fontsize{8}{5.5}
    \def\arraystretch{1.3}
    \setlength{\tabcolsep}{1.55pt}
    \setlength{\abovetopsep}{0pt}
    \setlength\belowbottomsep{0pt} 
    \setlength\aboverulesep{0pt} 
    \setlength\belowrulesep{0pt}
\caption{Universal representation evaluation on SUPERB.}
\vspace{-1.0em}
\scalebox{1}{
\begin{tabular}{l|c|cc|c|c}
\toprule
\multirow{2}{*}{\textbf{Methods}} 
& \multirow{2}{*}{\textbf{\#Param}}
&  \multicolumn{2}{c|}{\textbf{SE}} & \textbf{SS} & \textbf{SD} \\ 
\cmidrule{3-6}

&  & \textbf{PESQ$\uparrow$} & \textbf{STOI$\uparrow$} &\textbf{SI-SDRi$\uparrow$} & \textbf{DER$\downarrow$} \\
\midrule
FBANK & -- & 2.55 & 93.60 & 9.23 & 10.05 \\
\midrule
Wav2vec2.0 Base~\cite{hsu2021hubert} & 95.04M & 2.55 & 93.9 & 9.77 & 6.08 \\

HuBERT Base~\cite{hsu2021hubert} & 94.68M & 2.58 & 93.90 & 9.36 & 5.88 \\

Wang et.al~\cite{wang2023adapter} & 95.02M & -- & -- & 10.59 & -- \\ 


C-HuBERT Base~\cite{fazel2023cocktail} & 96.00M & \textbf{2.63} & 94.00 & 11.08 & 2.77 \\ \hline 


WavLM Base~\cite{chen2022wavlm} & 94.70M & 2.58 & 94.00 & 10.37 & 4.55 \\





SA-WavLM (Ours) & 103.7M & 2.62 & \textbf{94.18} & \textbf{11.13} & \textbf{1.88} \\

\toprule

\end{tabular}}
\label{tbl:superb}
\vspace{-1.0em}
\end{table}










\begin{table}[t]
\centering
    \fontsize{8}{6}\selectfont
    \def\arraystretch{1.8}
    \setlength{\tabcolsep}{4pt}
    \setlength{\abovetopsep}{0pt}
    \setlength\belowbottomsep{0pt} 
    \setlength\aboverulesep{0pt} 
    \setlength\belowrulesep{0pt}
\caption{Results on multi-speaker speech recognition.}
\vspace{-1.0em}
\begin{tabular}{l|c|cc}
\toprule
\multirow{2}{*}{\textbf{Pre-trained Model}} &  \multirow{2}{*}{\textbf{Spk. Embedding}} & \multicolumn{2}{c}{\textbf{WER (\%)\,}} \\ \cline{3-4}
& & \textbf{w/o LM} & \textbf{w/ LM} \\ 

\midrule

\multirow{2}{*}{HuBERT Base~\cite{hsu2021hubert}} & \xmark & 22.70  & 15.60 \\
 & \cmark & 17.42 & 14.88\\
\hline 

\multirow{2}{*}{WavLM Base~\cite{chen2022wavlm}} & \xmark & 15.97 & 10.38 \\
 & \cmark & 13.30 & 11.36 \\
\hline 


SA-WavLM (Ours) & \cmark & \textbf{8.39} & \textbf{6.49} \\

\toprule

\end{tabular}
\label{tbl:multi-speaker-asr}
\vspace{-2.0em}
\end{table}

\vspace{-0.2cm}
\subsection{Multi-speaker speech recognition}
\vspace{-0.1cm}
Following the utterance group-based evaluation settings in~\cite{huang2023adapting}, we evaluate SA-WavLM's performance on multi-speaker ASR. The character-level connectionist temporal classification (CTC) loss is used to fine-tune the pre-trained models except for the CNN encoder. We consider two settings: one without and one with speaker embeddings. In the former, permutation invariant training (PIT)~\cite{yu2017permutation} is used. In the latter, the same CLN operation as in SA-WavLM is applied to integrate speaker embeddings. 

Table~\ref{tbl:multi-speaker-asr} reports the word error rate (WER) for all the pre-trained models, with and without the 4-gram language model (LM). For models trained using PIT, concatenated minimum-permutation word error rate (cpWER)~\cite{watanabe20b_chime} is reported. As shown in Table~\ref{tbl:multi-speaker-asr}, WavLM, designed for mixture speech, obtains a much better performance compared to HuBERT which is designed for clean speech. When incorporating the speaker embeddings into the pre-trained models, both HuBERT and WavLM demonstrate reduced WER in settings without LM. Compared to WavLM, SA-WavLM achieves a more significant relative WER reduction of 37.4\%. This can be attributed to SA-WavLM's consideration of speaker interactions, which improves its speaker discrimination ability.



\begin{table}[t]
\centering
    \fontsize{8}{5.5}\selectfont
    \def\arraystretch{1.8}
    \setlength{\tabcolsep}{1.55pt}
    \setlength{\abovetopsep}{0pt}
    \setlength\belowbottomsep{0pt} 
    \setlength\aboverulesep{0pt} 
    \setlength\belowrulesep{0pt}
\caption{Results on speech extraction. The default stride for BSRNN is 8ms. We changed it to 5ms so that BSRNN's features can be aligned with the pre-trained features of 20ms stride.}
\vspace{-1.0em}
\scalebox{1}{
\begin{tabular}{l|c|c|cc}
\toprule
\textbf{Extraction Model}
& \textbf{Stride}
& \textbf{Pre-trained Model}
& \textbf{SDRi$\uparrow$}
& \textbf{SI-SDRi$\uparrow$} \\
\midrule
\multirow{5}{*}{\shortstack{BSRNN \cite{yu23b_interspeech}}} & 8ms & --  & 14.0  & 13.01 \\
\cmidrule{2-5}
& \multirow{4}{*}{5ms} & --  & 12.39  & 10.06 \\

& & HuBERT Base~\cite{hsu2021hubert} & 14.83  & 14.11 \\

& & WavLM Base~\cite{chen2022wavlm} & 14.64 & 15.47 \\



& & SA-WavLM (Ours) & \textbf{17.74}  & \textbf{17.3}  \\
\toprule

\end{tabular}}
\label{tbl:pSE}
\vspace{-0.8em}
\end{table}

\begin{table}[t]
\centering
    \scriptsize
    \fontsize{8}{5}\selectfont
    \def\arraystretch{1.8}
    \setlength{\tabcolsep}{2pt}
    \setlength{\abovetopsep}{0pt}
    \setlength\belowbottomsep{0pt} 
    \setlength\aboverulesep{0pt} 
    \setlength\belowrulesep{0pt}
\caption{Results on speech separation, including 1\%, 10\%, and 100\% of training data (13,900 utterances for 100\%).}
\vspace{-1.0em}
\scalebox{1}{
\begin{tabular}{l|c|c|cc}
\toprule
\textbf{Train}
& \textbf{Pre-trained} & \textbf{Separation}
& \multirow{2}{*}{\textbf{SDRi$\uparrow$}} 
& \multirow{2}{*}{\textbf{SI-SDRi$\uparrow$}}  \\
\textbf{Data}&\textbf{Model} &\textbf{Model} & & \\
\midrule
\multirow{4}{*}{1\%}&  --  &\multirow{4}{*}{ConvTasNet~\cite{luo2019conv}} & 3.05 & 2.59  \\

& HuBERT Base~\cite{hsu2021hubert}   & & 4.45 & 3.97  \\


& WavLM Base~\cite{chen2022wavlm} & & 7.56 &	7.09 \\



& SA-WavLM (Ours) & & \textbf{8.81} & \textbf{8.42} \\

\midrule

\multirow{4}{*}{10\%}&  --  &\multirow{4}{*}{ConvTasNet~\cite{luo2019conv}} & 9.73 & 9.16  \\

& HuBERT Base~\cite{hsu2021hubert}  & & 11.08 & 10.67  \\


& WavLM Base~\cite{chen2022wavlm} & & 11.93 & 11.58 \\



& SA-WavLM (Ours) & & \textbf{13.93} & \textbf{13.62} \\
\midrule

\multirow{4}{*}{100\%}&  --  &\multirow{4}{*}{ConvTasNet~\cite{luo2019conv}} & 14.53 & 14.12  \\

& HuBERT Base~\cite{hsu2021hubert}  &  & 14.99 & 14.62  \\


& WavLM Base~\cite{chen2022wavlm} & & 15.95 &	15.6 \\



& SA-WavLM (Ours) & & \textbf{16.55} & \textbf{16.22} \\
\toprule

\end{tabular}}
\label{tbl:low-resource}
\vspace{-1.9em}
\end{table}

\vspace{-0.1cm}
\subsection{Speech extraction and separation}
\vspace{-0.1cm}
While SUPERB offers a standard and comprehensive benchmark for the research community, many works have investigated to evaluate SSL models' capabilities 
in different ways, striving to push the performance boundaries further~\cite{chen2022large,baevski2020wav2vec,hsu2021hubert,fan21_interspeech,morais2022speech}. 
However, the ability of the pre-trained model on speech extraction and separation is relatively under-explored. In~\cite{shi2021discretization}, discretization is first performed on the representations obtained from the frozen pre-trained model and the clean waveform is resynthesized from the discrete tokens. In our experiment, we use the representations from all layers of the frozen pre-trained models directly, as discretization could potentially result in the loss of information. The pre-trained representations of different layers are weighted-sum, upsampled and then concatenated with the supervised model's encoded features before feeding into separation/extraction modules in these models. 

Table~\ref{tbl:pSE} shows the results for speech extraction using BSRNN~\cite{yu23b_interspeech} that is integrated with different pre-trained models. The results indicate that BSRNN benefits from integrating with the pre-trained speech representations. SA-WavLM stands out among all the pre-trained models, achieving 3.74dB SDRi and 4.29dB SI-SDRi improvements from BSRNN (8ms).

Table~\ref{tbl:low-resource} presents the results for speech separation using ConvTasNet~\cite{luo2019conv}. In this experiment, we further investigate the effectiveness of the pre-trained models in low-resource scenarios (e.g. 1\%). Consistent with the findings in Table~\ref{tbl:pSE}, ConvTasNet benefits from the pre-trained representations across all settings (i.e., 1\%, 10\% and 100\%). The benefits are particularly prominent in low-resource scenarios. This 
can be explained as pre-trained representations provide more noise-invariant contextual information. 
Again, among all the pre-trained models, SA-WavLM demonstrates the most promising performances. 

\vspace{-0.1cm}
\section{Conclusion}
\vspace{-0.1cm}
This paper proposes SA-WavLM, a novel speaker-aware self-supervised pre-trained model for mixture speech. Pre-trained with the designed ``extract-merge-predict'' pipeline, SA-WavLM considers the presence of all speakers in the mixture speech and models the interactions between them. In addition, a speaker shuffling strategy is introduced to ensure the model's invariance towards speaker order and absence. 
Experimental results show that SA-WavLM has an enhanced ability to remove interference across various mixture speech tasks. 


\newpage 
\bibliographystyle{IEEEtran}
\bibliography{mybib}

\end{document}